\documentclass[aps,pre,twocolumn,superscriptaddress,floatfix,longbibliography]{revtex4-2}

\usepackage{amsmath,amssymb}
\usepackage{graphicx}
\usepackage{booktabs}
\usepackage[colorlinks=true,linkcolor=blue,citecolor=blue,urlcolor=blue]{hyperref}

\begin{document}

\title{Absorbing phase transition in a queueing model\\
of coupled adaptive agents}

\author{Alexei Vazquez}
\email{alexei@nodeslinks.com}
\affiliation{Nodes \& Links Ltd, Salisbury House, Station Road,
Cambridge, CB1 2LA, UK}

\date{\today}

\begin{abstract}
What decides whether people do things together or separately? Many activities
cannot be carried out alone, and an individual must rank them against the
private tasks competing for the same time. We address this within the
priority-queue description of human activity by letting each agent \emph{choose}
the priority of a shared task rather than drawing it from a fixed distribution:
the value of the joint activity, discounted by the estimated risk that the
partner will not take part. Participation becomes strategic, and the model
acquires a phase transition. A coupled phase, in which joint activity is
sustained, is separated from an absorbing solitary phase by a saddle-node
bifurcation that we obtain in closed form. The transition is discontinuous and
the solitary phase is absorbing, so collapse is irreversible unless an agent
persists unilaterally for of order one memory time, a cost we also compute. The
heavy-tailed interevent statistics that motivate queueing models of human
dynamics survive only in a narrow window at the transition, and there the
exponent is fixed by the fraction of time spent coupled rather than by queue
length: the universality classes of the non-strategic model do not survive the
introduction of choice. On a network, attention divides as $1/(k+a)$ and fixes a
critical degree beyond which no coupled state exists, so the solitary phase
percolates according to the Molloy--Reed criterion with the second moment
truncated at that degree --- formally an attack on hubs, with no attacker. For
group activities the critical degree falls as the $(m-1)$th root, implying a
maximum group size. The transition organises two quantities already measured in
communication records: a finite capacity for keeping ties active, and the decay
of ties whose rhythm is interrupted.
\end{abstract}

\maketitle

\section{Introduction}

Human dynamics asks how people allocate their time among the things they might
do, and what the timing of the resulting activity reveals about the decision
process behind it. Its standard framework is queueing
theory~\cite{barabasi2005,vazquez2005,vazquez2006}: an individual's to-do list
is a priority queue executed highest-priority-first, which produces power-law
interevent time distributions $P(\tau)\sim\tau^{-\alpha}$, with $\alpha=1$ and
$\alpha=3/2$ identified as universality classes and corroborated by electronic
and postal correspondence data~\cite{barabasi2005,oliveira2005,vazquez2006},
the $\alpha=3/2$ class going back to the single-server analysis of
Cobham~\cite{cobham1954}.

The question we address is one this framework leaves open. A person's list is
not made only of things that can be done alone. A meeting, a conversation, a
shared meal happen only if somebody else chooses them at the same moment. How
should such an activity be ranked against the private tasks competing for the
same hour, and what becomes of the timing statistics when everyone ranks that
way?

The first half of the question has been posed. A minimal two-agent
extension~\cite{oliveira2009} gives each agent one \emph{interacting} task $I$,
executable only when both agents select it in the same step, and one aggregate
\emph{non-interacting} task $O$ standing for $L_j-1$ private tasks, predicting a
countable family of exponents, $\alpha=1+1/\max_j(L_j-1)$, interpolating between
$2$ and $1$: the more distracted agent sets the exponent.

\begin{figure*}[t]
\includegraphics[width=\textwidth]{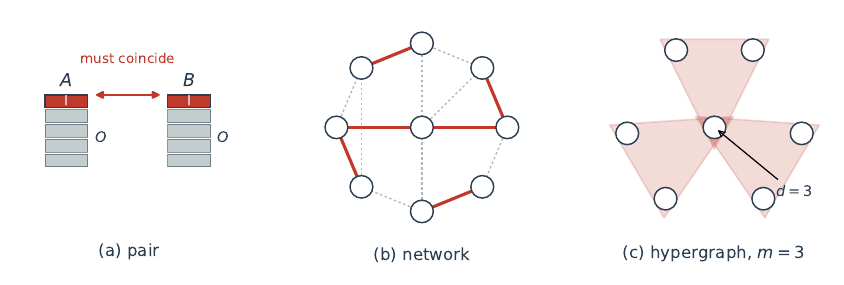}
\caption{The three settings. An agent holds one interacting task per shared
activity plus an aggregate private task $O$ standing for $L-1$ individual tasks,
and executes one item per step. (a) A pair: $I$ executes only if both select it.
(b) A network: one interacting task per neighbour, so attention is divided;
solid links are coupled relations, dashed ones have collapsed. (c) A hypergraph:
an activity of size $m$ executes only if all $m$ members select it, and an agent
may belong to $d$ groups.}
\label{fig:schematic}
\end{figure*}

In that treatment the priority of the interacting task is drawn independently
from a fixed density. This is the natural assumption if the shared activity is
exogenous, but it removes the feature that makes joint tasks distinctive: an
agent who is repeatedly stood up should stop offering. Here we make the
interacting-task priority a decision variable. The agents estimate each other's
participation rate from observation and set their priority accordingly. This
changes the object from a stochastic process into a coupled inference problem
and, as we show, changes its phenomenology substantially.

Once participation is chosen rather than imposed, the object of study changes.
The question is no longer the shape of one distribution but the collective state
of a set of agents, and the natural instrument for it is statistical mechanics.
Human dynamics already borrows from that apparatus at the level of exponents and
universality classes, and interaction is known from the theory of critical
phenomena to be decisive for universality~\cite{stanley1971}. We argue that here
the informative structure lies one level up. The model has two phases, an order
parameter that separates them, and a transition of definite character; the
exponents turn out to be a property of where the system sits relative to that
transition rather than the object of study. Phases and transitions, not
exponents, are what the model has to say about human activity.

Strategic queueing is of course an established subject, from
Naor~\cite{naor1969} onward, but it concerns customers deciding whether to join,
balk, or purchase priority in a service queue. What is at issue here is
different: the priority of a task that cannot be executed unilaterally, so that
the queueing model is simultaneously a coordination game with an absorbing
failure state. The resulting phenomenology
connects to two literatures with little contact so far with the queueing account
of human dynamics. The capacity constraint derived below has been measured
directly: mobile-phone records show a finite communication capacity limiting how
many ties an individual holds active, with larger networks maintaining weaker
ties rather than more communication~\cite{miritello2013,dunbar1992,saramaki2014}.
So has the timescale condition: ties interrupted for much longer than their
established rhythm decay~\cite{navarro2017}. And the percolation criterion we
obtain is formally that of intentional
attack~\cite{cohen2000,cohen2001,albert2000}, with no adversary present ---
attention scarcity removes high-degree relations by itself.

\section{Model}
\label{sec:model}

\subsection{Definition}

We state the model in full, so that it can serve as a reference point. It
retains the substrate of Ref.~\cite{oliveira2009} and adds rule (vi).

\begin{enumerate}
\item[(i)] \emph{Tasks.} Agent $j$ holds one aggregate private task $O$ standing
for $a_j\equiv L_j-1$ individual tasks, and one interacting task $I$ for each
shared activity it belongs to.
\item[(ii)] \emph{Priorities.} The private-task priority is the largest of $a_j$
independent uniform variates, of density $f_{O_j}(x)=a_j x^{a_j-1}$ on $[0,1]$,
redrawn whenever $O$ is executed. The interacting-task priority is $x_{I_j}$,
held until the task is acted on.
\item[(iii)] \emph{Selection.} Each agent executes the single highest-priority
item in its list per step.
\item[(iv)] \emph{Execution.} A shared activity executes only if every one of
its members selects it in the same step.
\item[(v)] \emph{Payoffs.} Each participant receives $R_I$ when the activity
executes; an agent whose offer is not reciprocated receives $x_{O_j}-c$, the
step being wasted; otherwise an agent receives $x_{O_j}$, the realised value of
its private task.
\item[(vi)] \emph{Policy.} $x_{I_j}$ is set by Eq.~\eqref{eq:policy} from an
estimate $\hat p$ of the probability that the others participate, maintained as
an exponential moving average with retention $\lambda=e^{-1/\tau_{\rm mem}}$
over observed selections. Independently, an agent offers regardless of belief
with probability $\epsilon$ per step.
\end{enumerate}

Rules (i)--(v) are those of Ref.~\cite{oliveira2009}, in which $x_{I_j}$ is
drawn from a fixed density; rule (vi) replaces that draw by a decision and is
the whole of what we add. The pair, network and hypergraph settings of
Fig.~\ref{fig:schematic} differ only in how many interacting tasks an agent
holds and how many members an activity requires. Two amendments are needed once
an agent holds more than one interacting task, and are given in
Sec.~\ref{sec:networks}.

The exploration rate $\epsilon$ in (vi) is not a numerical convenience. At
$\epsilon=0$ the state $p=0$ is strictly absorbing: an agent that stops
participating never observes its partner participating and therefore never
revises. Escape requires two agents to explore in the same step, at rate
$\epsilon^2$.

\subsection{Priority as a decision}

Since $O$ is drawn afresh each step, the probability that $j$ selects $I$ is
\begin{equation}
  p_j = \Pr\!\left(x_{O_j} < x_{I_j}\right) = x_{I_j}^{\,a_j},
  \label{eq:reachprob}
\end{equation}
so the agent controls a propensity, not a decision: with $a_j$ large, even a
high $x_{I_j}$ yields little participation.

The cost $c$ of an unreciprocated offer is what makes a model of the partner
necessary; at $c=0$ the optimal policy is $x_{I_j}=1$ whatever the partner does.

Writing $\hat p$ for the agent's estimate of its partner's participation
probability, selecting $I$ is preferable whenever
$\hat p R_I + (1-\hat p)(x_{O_j}-c) > x_{O_j}$, giving the optimal priority
\begin{equation}
  x_{I_j}^{*} = \min\!\left\{1,\;\max\!\left\{0,\;
              R_I - c\,\frac{1-\hat p}{\hat p}\right\}\right\},
  \label{eq:policy}
\end{equation}
clipped to the unit interval: \emph{the value of the interaction discounted by
the risk that the partner will not participate}.
Estimates are maintained by an exponential moving average with retention
$\lambda=e^{-1/\tau_{\rm mem}}$ over the partner's observed selections.

\subsection{Levels of modelling}

We distinguish three policies by what each agent represents.

\begin{description}
\item[Level 0] No partner model: $x_I = R_I$, clipped.
\item[Level 1] Partner model: $\hat p$ tracked, Eq.~\eqref{eq:policy} applied.
\item[Level 2] Self model: the agent additionally tracks $\hat s$, its estimate
  of \emph{the partner's estimate of itself}, and evaluates a finite-horizon
  rollout of the coupled dynamics in which its own participation raises $\hat s$,
  which raises the partner's priority through Eq.~\eqref{eq:policy}, which raises
  the partner's participation. It commits to $x_I=1$ when that rollout dominates
  the myopic choice.
\end{description}

Level 2 is the minimal policy requiring an agent to represent itself as an
object in another's model; the ladder is that of the cognitive-hierarchy and
level-$k$ literature~\cite{camerer2004}.

\section{Fixed points}

For a symmetric pair the participation probability obeys $p \mapsto
\mathcal{F}(p)$ with
\begin{equation}
  \mathcal{F}(p) = \Bigl(\min\bigl\{1,\max\{0,\,R_I - c(1-p)/p\}\bigr\}\Bigr)^{a}.
\end{equation}
Setting $\mathcal{F}(p)=p$ and requiring tangency gives the saddle-node in
closed form,
\begin{equation}
  p^{*} = (a c)^{a/(a+1)}, \qquad
  R_c = (p^{*})^{1/a} + \frac{c}{p^{*}} - c .
  \label{eq:rcrit}
\end{equation}
For $c=0.35$ this gives $R_c=0.8332$ at $p^*=0.5916$ for $L=2$, and
$R_c=0.9818$ at $p^*=0.7885$ for $L=3$.

\begin{figure*}[t]
\includegraphics[width=\textwidth]{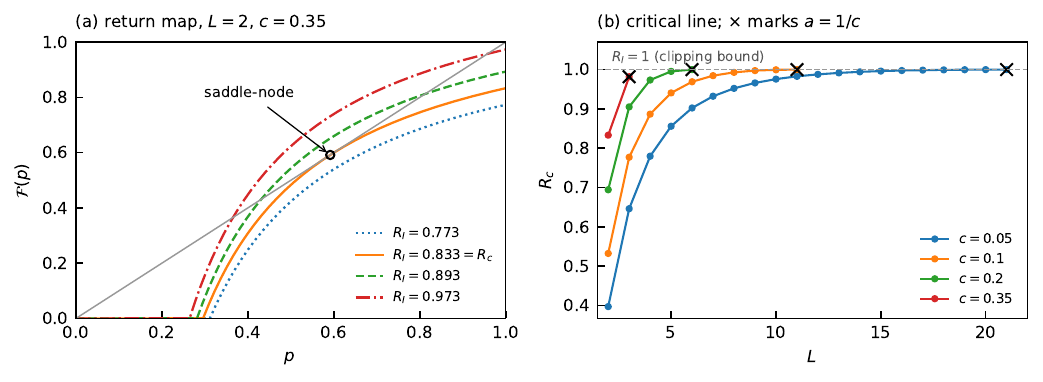}
\caption{(a) Return map $\mathcal{F}(p)$ for $L=2$, $c=0.35$, with the diagonal
in grey; the two nontrivial fixed points are born in a saddle-node at
$R_c=0.8332$, $p^*=0.5916$ (circle). (b) Critical line $R_c(L)$ from
Eq.~\eqref{eq:rcrit}; crosses mark the validity boundary $a=1/c$,
Eq.~\eqref{eq:validity}.}
\label{fig:phase}
\end{figure*}

Equation~\eqref{eq:rcrit} is meaningful only while $p^*\le 1$, that is
\begin{equation}
  a \le 1/c .
  \label{eq:validity}
\end{equation}
Above this bound the tangency lies outside the accessible range and there is no
interior bifurcation: the coupled phase is created discontinuously at the
clipping boundary $x_I=1$. At $c=0.35$ this restricts the continuous transition
to $L\le 3.86$. Figure~\ref{fig:phase} shows the return map and the critical
line.

Below $R_c$ only the solitary fixed point $p=0$ exists. Above it there are two
stable phases, coupled and solitary, separated by an unstable point whose
position rises with $a$: more private tasks enlarge the basin of solitude. The
structure is that of a stag hunt~\cite{skyrms2004}, whose behaviour on growing
and structured networks has been studied in its own right~\cite{starnini2011},
with the solitary phase risk-dominant and absorbing at $\epsilon=0$.

\section{Results}

Unless stated otherwise, simulations use $\tau_{\rm mem}=200$, symmetric agents,
and $\ge 10^5$ steps after a $20\%$ burn-in. We quantify the phase by the
\emph{occupancy} $\phi$, the fraction of time with $p>p^*/2$.

\subsection{Validation}

Reproducing Ref.~\cite{oliveira2009} with i.i.d.\ priorities recovers
$\alpha=1+1/\max_j(L_j-1)$. Asymmetric pairs sit on the prediction,
$(L_A,L_B)=(2,3)$ giving $\alpha=1.500$ and $(2,5)$ giving $1.250$, both stable
across four decades of the tail; symmetric pairs approach from below ($L=2$
gives $1.82$ at the $90$th percentile, drifting to $1.94$ deeper in), as the
logarithmic correction at $a_A=a_B$ requires. The mean interevent time diverges
with the observation window as $T^{2-\alpha}$, so for $\alpha\le2$ the
interaction has no characteristic timescale.

\subsection{Modelling the partner's model}

Level-2 agents remain coupled well below the level-1 threshold. At $L=3$ a
level-1 pair collapses to the solitary phase for every $R_I\le 0.99$, whereas a
level-2 pair remains coupled down to $R_I=0.90$. The effect is markedly
asymmetric in a way that inverts the $\max_j$ rule of the non-adaptive
model: at $L=3$ and $R_I\in\{0.94,0.96,0.98\}$, pairs of
level-1 agents collapse while a level-2/level-1 pair remains coupled
(Table~\ref{tab:asym}). In the non-adaptive model the more distracted agent sets
the exponent; in the adaptive model the more sophisticated agent sets the phase.

\begin{table}[b]
\caption{Asymmetric pairing at $L=3$, $\tau_{\rm mem}=50$. A single level-2
agent holds the pair in the coupled phase.}
\label{tab:asym}
\begin{ruledtabular}
\begin{tabular}{ccll}
$R_I$ & levels $(A,B)$ & phase & occupancy \\
\colrule
0.94 & (1,1) & solitary & 0.000 \\
0.94 & (2,1) & coupled  & 0.942 \\
0.96 & (1,1) & solitary & 0.000 \\
0.96 & (2,1) & coupled  & 0.961 \\
0.98 & (1,1) & solitary & 0.000 \\
0.98 & (2,1) & coupled  & 0.980 \\
\end{tabular}
\end{ruledtabular}
\end{table}

\subsection{Persistence across silence}

We next decouple the rate of \emph{opportunity} from willingness by allowing an
interaction to be possible only with probability $\nu$ per step, and by making
the partner's action unobservable except when an interaction occurs. The mean
gap between opportunities, $1/\nu$, then sets the rate at which information
about the partner arrives.

If unobserved steps decay the estimate --- silence treated as refusal --- the
coupled phase is lost whenever the estimate comes to be dominated by
accumulated silence rather than by the last observed encounter. At $1/\nu=100$,
pairs with $\tau_{\rm mem}\in\{5,50\}$ remain coupled while
$\tau_{\rm mem}\in\{500,5000\}$ collapse.

The decisive variable, however, is not the ratio of timescales but whether the
estimate is held at all. If unobserved steps leave $\hat p$ unchanged, coupling
survives at \emph{every} opportunity rate and every memory constant tested, with
the realised interaction rate tracking $\nu$ to three digits
(Table~\ref{tab:hold}). A partner model maintained through intervals containing
no information is what makes rare interaction viable.

\begin{table}[t]
\caption{Occupancy at $L=3$, $R_I=0.99$, level 2, with the partner's action
unobservable between interactions. Left: silence decays the estimate. Right:
the estimate is held.}
\label{tab:hold}
\begin{ruledtabular}
\begin{tabular}{ccccc}
 & \multicolumn{2}{c}{decaying} & \multicolumn{2}{c}{held} \\
$1/\nu$ & $\tau_{\rm mem}=50$ & $5000$ & $\tau_{\rm mem}=50$ & $5000$ \\
\colrule
10  & 0.843 & 0.000 & 1.000 & 1.000 \\
100 & 0.756 & 0.000 & 1.000 & 1.000 \\
333 & 0.755 & 0.000 & 1.000 & 1.000 \\
\end{tabular}
\end{ruledtabular}
\end{table}

\subsection{Loss of universality}

Away from $R_c$ the adaptive system is not scale-free: it polarises to
$\tau=1$ in the coupled phase and $\tau\to\infty$ in the solitary one, with
apparent tail indices in the range $4$--$40$. Heavy tails reappear only in a
narrow window above $R_c$, where the two fixed points nearly coincide and belief
noise drives intermittent switching. Inside that window the interevent
coefficient of variation reaches $8$--$50$ and the tail is favoured over an
exponential by log-likelihood differences of $10^{3}$ to $10^{5}$, using the
tail-comparison procedure of Ref.~\cite{clauset2009}.

The window exists only for $\epsilon>0$. At $\epsilon=0$ the solitary phase is
absorbing, the pair falls once and never returns, and no value of $R_I$ produces
a heavy tail.

Crucially, the exponent measured in the window is \emph{not} given by
$\alpha=1+1/\max_j(L_j-1)$. Because the transition is sharp, $\phi$ cannot be
tuned through $R_I$; varying only the noise realisation at a fixed operating
point, however, spreads $\phi$ over a wide range, and $\alpha$ tracks it
(Table~\ref{tab:occ}). The correlation between $\phi$ and $\alpha$ is $+0.70$ at
$L=2$ and $+0.76$ at $L=6$; the two $\alpha(\phi)$ curves nearly coincide,
while the non-adaptive prediction would require $2.00$ and $1.20$ respectively.

\begin{table}[b]
\caption{Tail index versus occupancy at fixed operating points, varying only the
random seed. Rows at $\phi=1$ have ${\rm CV}\simeq0.76$ and are not heavy-tailed;
their fitted indices are reported for completeness only.}
\label{tab:occ}
\begin{ruledtabular}
\begin{tabular}{lcc}
 & $L=2$, $c=0.35$ & $L=6$, $c=0.10$ \\
\colrule
$1+1/\max_j(L_j{-}1)$ & 2.00 & 1.20 \\
${\rm corr}(\phi,\alpha)$ & $+0.70$ & $+0.76$ \\
$\alpha$ at $\phi\simeq0.22$ & 1.79 & --- \\
$\alpha$ at $\phi\simeq0.68$ & 4.59 & 4.32 \\
$\alpha$ at $\phi=1.00$ & 4.62--4.65 & 4.37--4.42 \\
\end{tabular}
\end{ruledtabular}
\end{table}

We conclude that $\alpha$ is not an order parameter of the adaptive system. The
occupancy $\phi$ is the natural replacement: bounded, directly measured, and
monotone in $R_I$, $L$ and modelling level, with the scale-free interevent
statistics appearing as a symptom of intermediate $\phi$ rather than as its
definition.

\subsection{Bimodality and irreversibility}

At fixed parameters ($L=2$, $c=0.35$, $R_I=0.8428$, $\epsilon=10^{-2}$) the
outcome over $140$ realisations is strongly bimodal: $45$ end with $\phi<0.25$,
$93$ with $\phi>0.75$, and only $2$ in between.

The outcome is nevertheless not fixed by early noise. Classifying the final
phase from a trajectory prefix barely exceeds the majority-class baseline of
$0.679$: accuracies are $0.679$, $0.693$, $0.693$, $0.714$, $0.736$ and $0.836$
for prefixes of $1$, $3$, $10$, $30$, $60$ and $120\times10^3$ steps. The first
$10^4$ steps carry essentially no information. Collapse is a rare event
available throughout the trajectory rather than a bifurcation resolved at the
outset.

Collapse is, however, irreversible without intervention: none of the $45$
collapsed realisations recovered when left alone. Promoting agents to level 2 at
the midpoint recovers $35.6\%$ of them, whereas pairs of level-2 agents run from
$t=0$ on the same seeds never collapse at all (Table~\ref{tab:rescue}).
Promoting both agents is worth nothing over promoting one: recovery is driven by
a single agent participating against the advice of its own estimate, and the
partner need only be able to respond.

\subsection{Patience, not reachability}

The partial recovery in Table~\ref{tab:rescue} might indicate states from which
no policy in this family can return. It does not. As specified above, the
level-2 rollout evaluates a single step of unreciprocated participation. We
therefore introduce a commitment length $k$: the number of consecutive steps of
unilateral participation the agent evaluates, and latches onto if it decides to
proceed.

The decision is deterministic given beliefs, so the reachable region can be
mapped directly. Table~\ref{tab:noreturn} reports the smallest $\hat p$ at which
an agent still chooses to participate, taking $\hat s=\hat p$. At $k=1$ the
agent participates only if it already believes its partner participates
$\gtrsim 46\%$ of the time --- a condition a collapsed pair fails by
construction. Increasing $k$ alone is insufficient, and increasing the lookahead
$H$ alone changes nothing at all: recovery is flat at $48.5\%$ for
$H\in\{10,\ldots,1000\}$ at $k=1$. Only when a long commitment is combined with
a discount horizon $1/(1-\gamma)$ long enough to see past it does the threshold
collapse to zero, and then every collapsed realisation recovers.

\begin{table}[t]
\caption{Smallest belief $\hat p$ at which the agent still participates
(left), and measured recovery of the collapsed realisations (right), versus
commitment length $k$ and discount $\gamma$. Recovery is $48.5\%$ in every cell
but one.}
\label{tab:noreturn}
\begin{ruledtabular}
\begin{tabular}{ccccc}
 & \multicolumn{2}{c}{$\min \hat p$} & \multicolumn{2}{c}{recovered} \\
$k$ & $\gamma=0.97$ & $\gamma=0.999$ & $\gamma=0.97$ & $\gamma=0.999$ \\
\colrule
1    & 0.463 & 0.458 & $48.5\%$ & $48.5\%$ \\
10   & 0.458 & 0.453 & $48.5\%$ & $48.5\%$ \\
100  & 0.428 & 0.313 & $48.5\%$ & $48.5\%$ \\
1000 & 0.428 & $10^{-6}$ & $48.5\%$ & $100\%$ \\
\end{tabular}
\end{ruledtabular}
\end{table}

The residual $48.5\%$ is not policy-driven: it is the fraction of realisations
in which exploration lifts $\hat p$ above the participation threshold and the
agent latches. No state is beyond recovery; the cap measures the reluctance of a
short-horizon agent to persist without evidence. The baselines in
Tables~\ref{tab:rescue} and~\ref{tab:noreturn} come from runs of different
length with different collapsed subsets and are not directly comparable.

\subsection{Endogenous commitment}

The commitment length above is imposed. Letting the agent evaluate its rollout
at each candidate length and latch onto the argmax makes it a property of its
beliefs. The argmax is uninformative, pinning to the largest candidate since
further length is free once the partner participates, so we characterise the
policy by the minimum patience that beats not persisting,
\begin{equation}
  k_{\min}(\hat p) = \min\{\,k:\ \mathcal{R}(k) > \mathcal{R}(0)\,\},
\end{equation}
with $\mathcal{R}(k)$ the rollout value under a $k$-step commitment.

Two regularities emerge, both in units of the memory constant $\tau_{\rm mem}$.
First, $k_{\min}$ does not diverge as $\hat p \to 0$; it saturates, and bisecting
at $\hat p = 10^{-6}$ gives $k_{\min}/\tau_{\rm mem} = 0.740, 0.740, 0.740,
0.738$ for $\tau_{\rm mem} = 50, 100, 200, 400$. The price of recovery is
therefore a fixed multiple of a memory time and is independent of how
unfavourable the current estimate is. The partner begins to respond earlier, at
$0.347\,\tau_{\rm mem}$, where its myopic priority turns positive at
$s>c/(R_I+c)=0.293$; persistence must continue about twice that long before the
gain covers the commitment.

Second, whether the agent will pay that price is set by its horizon in the same
units: the smallest $H^{*} = 1/(1-\gamma)$ at which an agent with $\hat p =
10^{-6}$ still commits is $H^{*} \simeq 1.5\,\tau_{\rm mem}$ (measured $1.50$ at
$\tau_{\rm mem} = 100, 200, 400$; the value $2.00$ at $\tau_{\rm mem} = 25, 50$
is an upper bound set by the resolution of the $H$ ladder).

These predict the simulation. At $\tau_{\rm mem}=200$, so $H^{*}\simeq300$, an
agent choosing its own commitment length recovers $48.5\%$ of collapsed
realisations at $\gamma \le 0.995$ ($H \le 200$) and $100\%$ at $\gamma=0.999$
($H=1000$), while the fixed $k=1$ control stays at $48.5\%$ throughout. Judging recovery by interaction rate instead --- unbiased, since $\phi$ exceeds
$1/2$ whenever one agent is latched whatever its partner does --- moves the
baseline to $72.7\%$ but leaves the contrast unchanged, with the mean rate
rising from $0.368$ to $0.639$ and the partner's payoff from $0.642$ to $0.734$
against a solitary $\langle x_O\rangle=0.5$. The recoveries are genuine.

The agent that restores the interaction is not the one that gains most from it:
the committing agent earns $0.62$--$0.70$ against its partner's $0.68$--$0.76$,
the difference being the unreciprocated offers it absorbs.

\begin{table}[t]
\caption{Intervention on the $45$ realisations that collapsed as level-1 pairs,
with promotion at the midpoint of the run.}
\label{tab:rescue}
\begin{ruledtabular}
\begin{tabular}{lcc}
intervention & recovered & mean final $\phi$ \\
\colrule
none (control)            & $0.0\%$   & 0.020 \\
agent $A$ $\to$ level 2   & $35.6\%$  & 0.369 \\
both $\to$ level 2        & $35.6\%$  & 0.369 \\
level 2 from $t=0$        & $100\%$   & 1.000 \\
\end{tabular}
\end{ruledtabular}
\end{table}

\section{Percolation on networks}
\label{sec:networks}

The two-agent model extends to a network by giving agent $i$ one interacting
task per neighbour alongside its solitary aggregate, with a single item selected
per step. Attention then becomes a scarce resource: an agent with $k$ relations
leaves $k-1$ offers unanswered at every step, and each relation must compete
with the others as well as with the private queue. An edge executes, as before,
only when both endpoints select it.

Two amendments to rule (ii) are needed, since abandoning either produces an
artefact. First, the interacting-task priority must be \emph{drawn} from a
density the policy controls, $u\,x_{I_j}$ with $u$ uniform, rather than being a
deterministic function of $\hat p$; otherwise each agent locks onto its single
best partner, every other relation starves, and the coupled subgraph is forced
to be a perfect matching. Second, an \emph{offer} must consume the attempt
whether or not it is reciprocated. Redrawing only on execution freezes the
choice: $i$ offers to the same partner indefinitely, and if that partner's best
is not $i$ the pair deadlocks.

The topology is fixed here and only the state of each edge evolves, but the
coupled subgraph is itself an effective network that grows and fragments, so the
setting is adjacent to the literature on adaptive
networks~\cite{gross2008}, where the interplay of node dynamics with topology
generates discontinuous transitions and bistability of the kind seen below.

A capacity limit on the number of interactions a node may sustain has been
imposed before in a game-theoretic setting: restricting the associative capacity
of nodes in a prisoner's dilemma on scale-free networks changes the level of
cooperation and admits an optimal capacity~\cite{poncela2011}. The distinction
here is that no capacity is imposed. It follows from Eq.~\eqref{eq:select},
because an agent executing one task per step divides its attention as
$1/(k+a)$, and the resulting $k_c$ is a property of the queue rather than a
parameter. The game is also different --- coordination with an absorbing
failure state rather than a dilemma --- so the two capacities are not the same
quantity, but they act on the same variable.

We use three growth rules. Triadic closure $LS(\ell=1)$ is the undirected
local-search rule in which a newcomer links to a random node and to one of its
neighbours, closing a triangle at every step in the manner long argued to
organise social ties~\cite{granovetter1973}; growth by triadic closure
generates community structure spontaneously, and randomising the attachment
destroys it~\cite{bianconi2014}. Against it we set the degree-preserving
randomization of $LS$, and preferential attachment
$BA(m=2)$~\cite{barabasi1999} as a non-local control. Randomizing $LS$
preserves the degree sequence exactly, so any difference between the two is
attributable to local structure alone.

We use $c$, the cost of an unreciprocated offer, as the control parameter, and
call an edge coupled when its interaction rate exceeds ten times the solitary
baseline $\epsilon^2$. The order parameter $S$ is the largest connected
component of the coupled subgraph as a fraction of $n$
(Table~\ref{tab:percolation}).

\begin{figure*}[t]
\includegraphics[width=\textwidth]{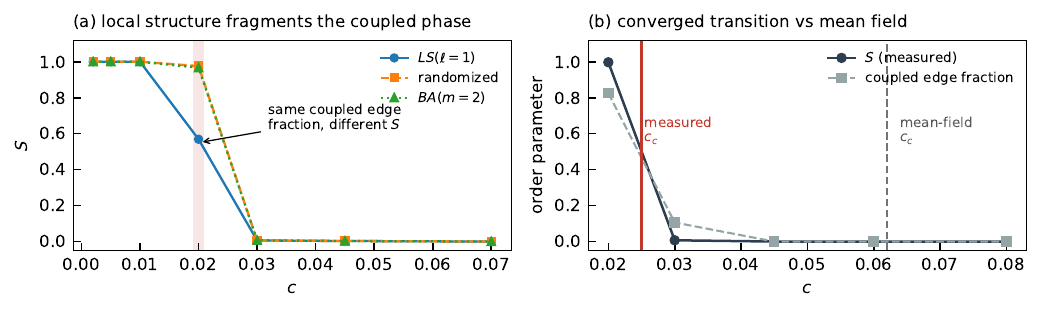}
\caption{Percolation of the solitary phase. (a) Giant component $S$ of the
coupled subgraph against the cost $c$ of an unreciprocated offer
($n=600$, $\langle k\rangle=4$, three realisations, $4\times10^4$ steps). At
$c=0.020$ all three carry the same coupled edge fraction but only the local rule
fragments. (b) Converged runs ($4\times10^5$ steps) place the transition between
$c=0.020$ and $c=0.030$; Eq.~\eqref{eq:cavity} predicts $c_c\simeq0.062$.}
\label{fig:percolation}
\end{figure*}

\begin{table}[t]
\caption{Coupled edge fraction and giant component $S$ of the coupled subgraph,
at $R_I=0.95$, $n=600$, $\langle k\rangle = 4$, averaged over three
realisations.}
\label{tab:percolation}
\begin{ruledtabular}
\begin{tabular}{lcccccc}
 & \multicolumn{2}{c}{$LS(\ell=1)$} & \multicolumn{2}{c}{randomized}
 & \multicolumn{2}{c}{$BA(m=2)$} \\
$c$ & edges & $S$ & edges & $S$ & edges & $S$ \\
\colrule
0.002 & 0.988 & 1.000 & 0.990 & 1.000 & 0.940 & 1.000 \\
0.010 & 0.950 & 1.000 & 0.955 & 1.000 & 0.874 & 1.000 \\
0.020 & 0.686 & \textbf{0.569} & 0.689 & \textbf{0.976} & 0.605 & 0.967 \\
0.030 & 0.183 & 0.006 & 0.178 & 0.006 & 0.165 & 0.007 \\
0.070 & 0.000 & 0.000 & 0.000 & 0.000 & 0.000 & 0.000 \\
\end{tabular}
\end{ruledtabular}
\end{table}

The solitary phase percolates between $c=0.02$ and $c=0.03$, and the
interesting behaviour is at the lower edge. There the two networks have
statistically identical coupled edge fractions, $0.686$ against $0.689$, and
identical degree-resolved survival, yet entirely different connectivity: the
triadic-closure network has $S=0.569$, thirteen components of size three or
more and a second of $99$ nodes, while its randomization is essentially intact
at $S=0.976$. Local structure does not determine which relations survive; it
determines how their failure is arranged. Under triadic closure the coupled phase fragments into separated
clusters, whereas at identical degrees without local structure it thins while
staying connected. The split between local and non-local rules that governs the
emergence of communities~\cite{bianconi2014} thus reappears as a split in the
way the solitary phase invades.

Survival falls monotonically with degree --- the fraction of a node's relations
that remain coupled at $c=0.020$ is $0.856$, $0.704$, $0.539$ and $0.346$ for
degrees near $2$, $4$, $8$ and $16$ --- and the two networks agree to within
$0.01$ at every degree. This follows directly from the attention constraint:
$\hat p \simeq P(\text{active})/k$, so a relation is sustainable only while
$k \lesssim P(\text{active})(R_I+c)/c$. High degree is a liability rather than a
protection, because what an agent competes for is its partner's attention, and
the coincidence of the two curves shows the effect to be independent of
topology.

\subsection{Analytic percolation threshold}

The transition admits a closed-form treatment. Since priorities are $uX$ with
$u\sim U(0,1)$ drawn per offer and the solitary task has $F_O(t)=t^{a}$, the
probability that an agent of degree $k$ selects one particular neighbour is
\begin{equation}
  \int_0^1 u^{k-1}(uX)^a\,du = \frac{X^{a}}{k+a},
  \label{eq:select}
\end{equation}
so attention divides as $1/(k+a)$. Self-consistency $X=\Psi(X^a/(k+a))$ with
$\Psi(p)=B-c/p$ and $B=R_I+c$ gives $X^{a+1}-BX^{a}+c(k+a)=0$, whose tangency at
$X^{*}=aB/(a+1)$ fixes a critical degree
\begin{equation}
  k_c=\frac{a^{a}(R_I+c)^{a+1}}{c\,(a+1)^{a+1}}-a ,
  \label{eq:kc}
\end{equation}
equal to $(R_I+c)^2/4c-1$ for $a=1$. Beyond $k_c$ no coupled state exists at any
belief: attention is divided $k$ ways, $\hat p \sim 1/k$, and it falls below the
collapse threshold $c/(R_I+c)$.

An edge is live only if sustained from both ends, so the process is
degree-targeted site percolation~\cite{stauffer1994} and the Molloy--Reed
criterion applies with the
second moment truncated at $k_c$,
\begin{equation}
  \sum_{k\le k_c} k(k-1)\,p_k > \langle k\rangle .
  \label{eq:mr}
\end{equation}
Equation~\eqref{eq:mr} uses the full degree, but a node whose relations have
already failed divides its attention only among the survivors. Writing $\phi$ for
the probability that an edge is sustained from one end and $q_k=kp_k/\langle
k\rangle$, the cavity equation closes on a locally tree-like graph as
\begin{equation}
  \phi=\sum_k q_k\,\mathbb{P}\!\left[\mathrm{Bin}(k-1,\phi)\le k_c-1\right],
  \label{eq:cavity}
\end{equation}
with a giant component iff $b=\sum_k q_k\,\mathbb{E}[(m-1)\mathbf{1}\{m\le
k_c\}]/\phi>1$, $m=1+\mathrm{Bin}(k-1,\phi)$. Setting $\phi=1$ recovers
\eqref{eq:mr}, which is therefore an upper bound. Exploration at rate $\epsilon$
sets $x=1$ and hence wins the selection, consuming the attention slot with
probability $1-(1-\epsilon)^k$; including it replaces $c(k+a)$ by
$c(k+a)(1-\epsilon)^{-k}$ in \eqref{eq:kc}.

Equation~\eqref{eq:kc} reproduces the measured degree dependence: survival falls
monotonically with $k$, and $k_c=10.8$ at $c=0.020$ sits where the measured
survival halves. The threshold itself, however, is a mean-field upper bound. On
the randomized network, where the tree-like assumption holds,
Eq.~\eqref{eq:cavity} gives $c_c\simeq0.062$ against a measured $c_c\simeq0.025$
--- long runs place the transition sharply between $c=0.020$ (coupled fraction
$0.825$, $S=0.997$) and $c=0.030$ ($0.106$, $S=0.007$). The discrepancy is the
expected one: a mean-field treatment cannot capture fluctuation-driven escape
into an absorbing state, and that escape is precisely the stag-hunt mechanism,
the runaway in which a falling $\hat p$ lowers $X$, which reduces the offers
received, which lowers $\hat p$ again. Relations that \eqref{eq:kc} declares
viable are destroyed by it.

Two points follow. Convergence near the transition is slow and directional:
over $2\times10^4$ to $3.2\times10^5$ steps the coupled fraction \emph{rises}
from $0.488$ to $0.802$ at $c=0.020$ while \emph{falling} from $0.165$ to
$0.082$ at $c=0.030$, so thresholds from short runs are biased toward the
initial condition. And the truncation in \eqref{eq:mr} acts at the \emph{top}
of the degree distribution. For $p_k\sim k^{-\gamma}$ with
$2<\gamma<3$ the second moment diverges and the network is robust to random
removal, but the removal here is targeted at high degree --- the one attack to
which such networks are most fragile. A scale-free contact structure is
therefore maximally vulnerable to this form of collapse, for a mechanical rather
than a topological reason: what an agent competes for is its partner's
attention, so high degree is a liability. The mathematics is that of intentional
attack~\cite{cohen2001}, but the attack is self-generated: no adversary removes
hubs, and the familiar contrast between robustness to failure and fragility to
targeted attack acquires an endogenous form.

The overestimate is also the expected behaviour of the approximation rather than
a defect peculiar to this model. Absorbing-state transitions in noisy spatial
coordination games belong to the directed-percolation class, whose upper
critical dimension is four~\cite{hinrichsen2000}, and mean-field treatments
systematically overestimate the extent of the ordered phase below it. We have
measured no exponents and make no claim of membership in that class.

We report the transition location but claim no exponent: $n=600$ with three
realisations is too small, and the agents here are all level 1, the network
analogue of the self-model being left for future work.

\section{Groups and hypergraphs}
\label{sec:groups}

Many joint activities are not pairwise. A meeting takes place only if everybody
turns up, which makes the interacting task a hyperedge rather than an edge.
Structures encoding interactions among three or more units, and the dynamics
they support, are reviewed in Ref.~\cite{battiston2020}. Encoding group
interactions on hypergraphs is by now a standard device for evolutionary games
in particular, where it changes the replicator dynamics and supplies a mechanism
for the survival of cooperation~\cite{alvarez2021,perc2013}; what follows
applies it to a coordination requirement inside a queue. We
therefore let agent $i$ hold one interacting task per hyperedge it belongs to,
write $d_i$ for that number, and require a hyperedge $e$ of size $m$ to execute
only when all $m$ of its members select it in the same step. Everything else is
as before.

Selection is unchanged, so Eq.~\eqref{eq:select} still gives $X^{a}/(d+a)$, but
an agent now needs the other $m-1$ members rather than one, so
$\hat p = [X^{a}/(d+a)]^{m-1}$ and self-consistency yields
\begin{equation}
  X^{n+1}-BX^{n}+c\,(d+a)^{m-1}=0,\qquad n=a(m-1),
  \label{eq:hyper}
\end{equation}
with $B=R_I+c$. This has the same form as the pairwise case with $a$ replaced by
$n$, so the tangency argument carries over and gives a critical group degree
\begin{equation}
  d_c(m)=\left[\frac{n^{n}B^{\,n+1}}{c\,(n+1)^{n+1}}\right]^{1/(m-1)}-a ,
  \label{eq:dc}
\end{equation}
which reduces to Eq.~\eqref{eq:kc} at $m=2$. The exponent $1/(m-1)$ carries the
content: the coordination requirement enters multiplicatively in the number of
people who must coincide, so $d_c$ falls very steeply with group size. At
$R_I=0.95$ it drops from $10.76$ to $1.60$ to $0.67$ for $m=2,3,4$ at $c=0.020$,
and from $112$ to $6.99$ to $2.51$ at $c=0.002$.

Because $d_c$ can fall below one, Eq.~\eqref{eq:dc} implies a \emph{maximum
group size} even for an agent that belongs to nothing else. Requiring
$d_c(m)\ge1$ gives $m_{\max}=2,3,4,5$ at $c=0.050,0.020,0.010,0.002$. It is
worth noting that freely forming conversational groups are observed to cap at
about four participants and to fission beyond five~\cite{dunbar1995}. Two
capacity arguments have been offered for that limit, one acoustic and
attentional~\cite{dunbar1995} and one based on a bound on the number of other
minds a participant can model at once~\cite{krems2016}. The present mechanism is
a third of the same character --- attention as a scarce resource in a queue ---
and we make no claim that it is the operative one.

Simulation confirms Eq.~\eqref{eq:dc} at $m=2$: the fraction of groups that
remain coupled falls from $1.000$ at $d=1$ to $0.309$ at $d=16$, straddling
$d_c=10.76$. At $m=3$ it fails. Equation~\eqref{eq:dc} gives $d_c=1.60$ and so
permits $d=1$, and the fixed point is genuinely there --- solving
\eqref{eq:hyper} at $m=3$, $d=1$, $c=0.020$ gives a stable root $X=0.862$
against an unstable one at $X=0.363$, and the simulation begins exactly on it,
with a measured meeting rate of $0.14$ against the predicted
$(X/2)^3=0.11$. Within a few thousand steps it nevertheless collapses to zero.

The reason is that a group fails if \emph{any single member's} belief drifts
below threshold, whereupon that member withdraws, the remaining members observe
no meeting, and their estimates follow. Escape from the coupled state is
therefore available through $m$ independent routes rather than one, while the
coordination probability itself falls as the $m$th power. Both effects push the
same way, and neither is visible to a mean-field treatment that asks only
whether a fixed point exists. The diagnosis is confirmed by suppressing the
noise: lengthening $\tau_{\rm mem}$, which reduces the variance of the belief
estimate, restores a nonzero meeting rate ($0$ at $\tau_{\rm mem}=200$ rising to
$0.006$ at $2\times10^4$ for $c=0.020$, and surviving already at
$\tau_{\rm mem}=10^3$ for $c=0.005$, where $d_c=4.08$ leaves a wider margin).

The rescue is only partial --- the recovered rates remain an order of magnitude
below the mean-field value --- so the conclusion is qualitative rather than
quantitative. The $\sim2.3$ discrepancy found in the pairwise network case
becomes, for $m\ge3$, a failure of kind: Eq.~\eqref{eq:dc} predicts viable
groups that the dynamics destroys. Group coordination is fragile not because
coordinated states fail to exist, but because the absorbing state is reachable
through any one participant.

\subsection{Reward proportional to group size}

A joint activity need not be worth the same however many take part: often the
value of the occasion grows with the number present. Replacing $R_I$ by
$R_I(m)=r\,m$ asks whether a reward growing with the group can offset a
coordination requirement that tightens with it.

The interior tangency disappears. Since $rm>1$ already at $m=2$ for $r$ of order
unity, the priority saturates at the clipping bound $x_I^{*}=1$ and
Eq.~\eqref{eq:dc} no longer applies: the coupled state is created at the
boundary, not in a saddle-node. Viability is the condition that the bound be
attained,
$c(1-\hat p)/\hat p \le R_I(m)-1$ with $\hat p=[1/(d+a)]^{m-1}$, giving
\begin{equation}
  d_c(m)=\left[\frac{r m-1+c}{c}\right]^{1/(m-1)}-a .
  \label{eq:dclin}
\end{equation}
A numerical solution of the full fixed-point map agrees with
Eq.~\eqref{eq:dclin} exactly, returning $X^{*}=1$ for every $m$ it admits.

This raises the ceiling substantially: at $r=0.95$ the maximum group size for a
singly-committed agent rises from $2,3,4,5$ to $8,9,10,13$ at
$c=0.050,0.020,0.010,0.002$. It does not remove it. Since
$(d+a)^{m-1}\le(R_I(m)-1+c)/c$ pits an exponential against a linear function,
$d_c\to0$ as $m$ grows, and arbitrarily large groups would need
$R_I(m)\gtrsim c\,(1+a)^{m-1}$ --- exponential in group size. No polynomial
increase in the value of an occasion compensates for the cost of getting
everybody free at once.

The simulations show a smaller gain still, and the discrepancy is instructive.
Equation~\eqref{eq:select} was obtained assuming that the priority of each task
is drawn afresh at every step. In the queueing substrate it is not: a priority
is redrawn only when its task is acted on, so a membership that is seldom
selected is seldom refreshed and its low priority persists. This is the same
mechanism that generates heavy-tailed waiting times in the original
model~\cite{barabasi2005,oliveira2009}, and here it depresses participation.
Balancing the flux with a redraw rate $u+\epsilon$ gives a stationary density
$\pi(u)\propto(u+\epsilon)^{-1}$, so a singly-committed agent participates with
probability
\begin{equation}
  q=\frac{1-\epsilon\ln\!\left[(1+\epsilon)/\epsilon\right]}
         {\ln\!\left[(1+\epsilon)/\epsilon\right]} ,
  \label{eq:qstat}
\end{equation}
equal to $0.207$ at $\epsilon=10^{-2}$ rather than the $1/2$ that a freshly
drawn priority would give. The measured meeting rate at $m=2$, $d=1$ is $0.0465$
against $q^2=0.0427$, where the fresh-priority value $1/4$ is wrong by a factor
of five. Replacing $\hat p$ by $q^{m-1}$ in the viability condition moves the
ceiling to $m_{\max}=4$, and the simulations have $m=2$ robust, $m=3$ marginal
at a rate of $10^{-3}$, and $m\ge4$ extinct.

The correction matters here because a singly-committed agent has one priority to
refresh; an agent holding many selects the largest, which is refreshed often, so
Eq.~\eqref{eq:select} is accurate where it was used above and worst in precisely
this case. The direction is unaffected: a reward proportional to group size
raises the largest sustainable group but does not abolish the limit, and raises
it by less than the existence criterion promises. What bounds the group is not
the worth of the occasion but the rate at which everybody is free at once.

\section{Conclusions}

Letting agents choose the priority of a task that requires someone else turns a
stochastic process into a system with a phase transition, and the transition
organises everything else.

\emph{Its nature.} The coupled and solitary phases are separated by a
saddle-node bifurcation, Eqs.~\eqref{eq:rcrit}, \eqref{eq:kc} and
\eqref{eq:dc}, obtained in closed form in all three settings. It is
discontinuous --- at $R_c$ a pair of fixed points is created at finite
participation --- and the solitary phase is absorbing, since an agent that stops
offering never observes one and so never revises. Together these give the asymmetry that is the
paper's practical content: the conditions that sustain joint activity are not
the conditions that restore it. None of our collapsed pairs recovered
spontaneously; recovery required unilateral persistence for
$0.74\,\tau_{\rm mem}$, and only an agent valuing the future beyond
$1.5\,\tau_{\rm mem}$ adopts it. A relation is cheap to keep and expensive to
rebuild, and the price is set by how long it takes to be forgotten rather than
by how unfavourable the present looks.

\emph{What it costs the established picture.} Heavy-tailed interevent times are
not generic once participation is chosen. Away from the transition the dynamics
polarises to $\tau=1$ or $\tau\to\infty$; scale-free statistics survive only in
a narrow window near the bifurcation, and only if agents retain a nonzero rate
of uninformed exploration, without which no window exists at all. Within the window the exponent tracks the
fraction of time spent coupled rather than the queue lengths, so the family
$\alpha=1+1/\max_j(L_j-1)$ does not survive. We propose phase occupancy,
bounded and directly measurable, in its place. Read the other way, exponents
close to that family in correspondence data are evidence that the priority
assignment there is close to exogenous.

\emph{On networks and in groups.} Attention divides as $1/(k+a)$, fixing a
critical degree $k_c$ beyond which no coupled state exists, and the solitary
phase percolates by Molloy--Reed with the second moment truncated at $k_c$ ---
the intentional-attack calculation~\cite{cohen2001} with no attacker. High
degree is a liability, because what one competes for is the partner's attention,
so a scale-free contact structure is maximally exposed to a collapse it
generates internally. Local structure does not change \emph{which} relations
survive, but it changes how their failure is arranged, fragmenting the coupled
phase into clusters rather than thinning it uniformly. For activities requiring
$m$ participants the critical degree falls as the $(m-1)$th root, giving a
maximum group size that a reward proportional to $m$ raises but cannot remove.

\emph{Relevance for human dynamics.} The transition makes contact with
quantities that have been measured, rather than only with exponents. A finite
capacity for keeping ties active is $k_c$, observed directly in communication
records~\cite{miritello2013,dunbar1992,saramaki2014}; here it is derived rather
than posited. The decay of ties whose rhythm is interrupted beyond a
characteristic multiple of their own frequency~\cite{navarro2017} is the
empirical form of our memory-timescale condition, and the most promising
quantitative test. Both are statements about the coupled phase and its boundary
rather than about the tail of an interevent distribution, which suggests that
the phase structure, not the exponent, is what the data have been describing.

\emph{Provenance and limitations.} Several results were reached while examining
a different hypothesis, that experience is a property of systems whose preferred
states are reachable only through another system they do not fully control;
nothing in Secs.~\ref{sec:model}--\ref{sec:groups} depends on it. The
critical points are mean-field: the percolation threshold exceeds the measured
one by a factor $\simeq2.3$, and for $m\ge3$ it fails qualitatively, predicting
viable groups that fluctuations destroy. Both are the expected behaviour of a
mean-field treatment of an absorbing-state transition~\cite{hinrichsen2000},
but we measure no exponents and claim no universality class. Convergence near
the transition is slow and directional. Finally, the commitment length is chosen
from a fixed menu, the recovery constants use a memory time that agents
attribute to their partners, and the network analogue of the partner-modelling
agent is not implemented.

\begin{acknowledgments}
All code is available at \url{https://github.com/av2atgh/interagents}.
The calculations were assisted by Claude Opus~5, an AI assistant developed by
Anthropic. Nodes \& Links Ltd provided support in the form of salary for Alexei
Vazquez but did not have any additional role in the conceptualization of the
study, the analysis, the decision to publish or the preparation of the
manuscript.
\end{acknowledgments}

\end{document}